\documentclass[12pt]{article}

\usepackage{scicite}
\usepackage{graphicx}
\usepackage{times}
\newcommand{\chil}{\chi_{\scriptscriptstyle L}}
\newcommand{\mub}{\mu_{\scriptscriptstyle B}}
\newcommand{\muq}{\mu_{\scriptscriptstyle Q}}
\newcommand{\mus}{\mu_{\scriptscriptstyle S}}
\newcommand{\nls}[1]{\chi_{\scriptscriptstyle B}^{(#1)}}
\newcommand{\snn}{\sqrt{s_{\scriptscriptstyle NN}}}
\newcommand{\tc}{T_c}

\newcommand{\etal}{{\sl et al.\/}}
\newcommand{\ie}{{\sl i.e.\/}}

\newcommand{\nat}{{\sl Nature\/}}
\newcommand{\pr}{{\sl Phys.\ Rev.\/}}
\newcommand{\prl}{{\sl Phys.\ Rev.\ Lett.\/}}

\newcommand{\np}{{\sl Nucl.\ Phys.\/}}
\newcommand{\pl}{{\sl Phys.\ Lett.\/}}
\newcommand{\rmp}{{\sl Rev.\ Mod.\ Phys.\/}}
\newcommand{\pt}{{\sl Phys.\ Today\/}}
\newcommand{\aip}{{\sl AIP\ Conf.\ Proc.\/}}

\newenvironment{sciabstract}{%
\begin{quote} \bf}
{\end{quote}}

\newcounter{lastnote}
\newenvironment{scilastnote}{%
\setcounter{lastnote}{\value{enumiv}}%
\addtocounter{lastnote}{+1}%
\begin{list}%
{\arabic{lastnote}.}
{\setlength{\leftmargin}{.22in}}
{\setlength{\labelsep}{.5em}}}
{\end{list}}

\title{
Scale for the Phase Diagram \\ of Quantum Chromodynamics
}

\author
{Sourendu Gupta,$^{1}$ Xiaofeng Luo,$^{2,3}$ Bedangadas Mohanty,$^{4}$ \\ Hans Georg Ritter,$^{3}$ Nu Xu$^{5,3}$\\
\\
\normalsize{$^{1}$Department of Theoretical Physics, Tata Institute of Fundamental Research,}\\
\normalsize{Homi Bhabha Road, Mumbai 400005, India}\\
\normalsize{$^{2}$Department of Modern Physics, University of Science and Technology of China,}\\
\normalsize{Hefei 230026, China}\\
\normalsize{$^{3}$Nuclear Science Division, Lawrence Berkeley National Laboratory, Berkeley,}
\normalsize{CA 94720, USA}\\
\normalsize{$^{4}$Experimental High Energy Physics and Applications Group, Variable Energy Cyclotron Centre,}\\
\normalsize{1/AF Bidhan Nagar, Kolkata 700064, India}\\
\normalsize{$^{5}$College of Physical Science and Technology, Central China Normal University,} \\
\normalsize{Wuhan 430079, China}\\
\\
}

\date{}

\begin{document}


\baselineskip24pt


\maketitle


\begin{sciabstract}

Matter described by Quantum Chromodynamics (QCD), the theory of strong
interactions, may undergo phase transitions when its temperature 
and the chemical potentials are varied. 
QCD at finite temperature is studied in the laboratory by colliding
heavy-ions at varying beam energies. We present a test of QCD in the
non-perturbative domain through a comparison of
thermodynamic fluctuations predicted in lattice computations with
the experimental data of baryon number distributions in high-energy heavy-ion collisions.
This study provides evidence for thermalization in these collisions, and 
allows us to find the crossover temperature between normal nuclear matter and
a deconfined phase called the quark gluon plasma. This value allows
us to set a scale for the phase diagram of QCD.

\end{sciabstract}

QCD is the theory of strong interactions---
one of the four fundamental interactions occurring in nature, and
an essential part of the standard model of particle physics. It
describes interactions between quarks and gluons, which are the
ultimate constituents of the majority of the visible mass of the
universe \cite{wilczek1,wilczek2}. In the short-distance regime where 
the momentum exchange between quarks and gluons is large, the strong 
coupling constant becomes small through the mechanism of asymptotic 
freedom. In this perturbative region QCD is very successful in explaining 
various processes observed in experiments involving electron-positron,
proton-proton and proton-antiproton collisions \cite{perturbation}.
In the non-perturbative regime tests of the theory were related to
the computation of hadron properties \cite{bmw}. In other regimes of
long-distance non-perturbative physics, the theory is yet to be tested.  
Here, we test the thermodynamics of bulk strongly interacting matter.

Experimental tests of non-perturbative QCD in the bulk can be carried out
by colliding heavy-ions (like U, Pb, Au, and Cu) at different 
center of mass energies, $\snn$ \cite{white1,white2,white3,white4}.  
Several experimental programs have
been launched or are in the planning stage at facilities such as the Large
Hadron Collider (LHC), the Relativistic Heavy Ion Collider (RHIC), the Super
Proton Synchrotron (SPS), the Facility for Anti-proton and Ion Research
(FAIR) and the Nuclotron-based Ion Collider fAcility (NICA), where 
the essential features of the QCD phase diagram can be studied. 

In QCD there are conserved quantities like the net-baryon number, $B$,
the net-electric charge, $Q$, and the net-strangeness, $S$. The term
net means the algebraic sum of the quantum numbers, where those of
anti-particles are the negatives of the corresponding particles.
As a result the thermodynamics of the bulk can be characterized by
the corresponding chemical potentials (energy needed to add/remove 
one unit of the conserved quantity to/from the system) $\mub$, $\muq$, and $\mus$ in
addition to the temperature, $T$, conjugate to the conserved energy of
a bulk system.  In experimental studies of particle ratios measured in
heavy-ion collisions it is observed that the relevant values of $\muq$
and $\mus$ are small compared to $\mub$.  
For example, in Au ion collisions within rapidity range of $\pm$ 0.1 unit 
at $\snn=200$ GeV (with impact parameter less than 3 fm)  one finds that 
$\mub=22\pm4.5$ MeV, while $\mus=3.9\pm2.6$ MeV and $\muq$ is still smaller \cite{starprc}. 

The lattice formulation of QCD is a non-perturbative approach
from first principles for obtaining the predictions of QCD.
Space-time is replaced by a lattice; quarks occupy the sites, and gluons occupy the links
between the sites. The lattice spacing, $a$, is the inverse of the cutoff
required to regulate any interacting quantum field theory. The theory is
solved numerically at several values of $a$. The
extrapolation to the continuum ($a=0$) can then be made through
the renormalization group equations. In QCD there is a conventional
temperature, $\tc$, which is an intrinsic scale of bulk hadronic
matter. We follow the definition that it is the temperature at
the peak of a susceptibility related to the confinement-deconfinement
order parameter (called the Polyakov loop susceptibility,
$\chil$) at $\mub=0$ \cite{lattice1,lattice2,lattice3,lattice4}. Lattice QCD computations show that 
this peak is finite, which corresponds to a cross-over \cite{crossover1,crossover2}. 
The temperature at which $\chil$ peaks, of course, changes with
$\mub$. However, once $\tc$ is known, such shifts as a function of $\mub$ 
can be quantified. This is similar to saying that the Celsius scale of temperature is defined
by the boiling point of water at normal pressure, $P$, and that the boiling point changes with $P$.

\begin{figure}
\begin{center}
\includegraphics[scale=0.35]{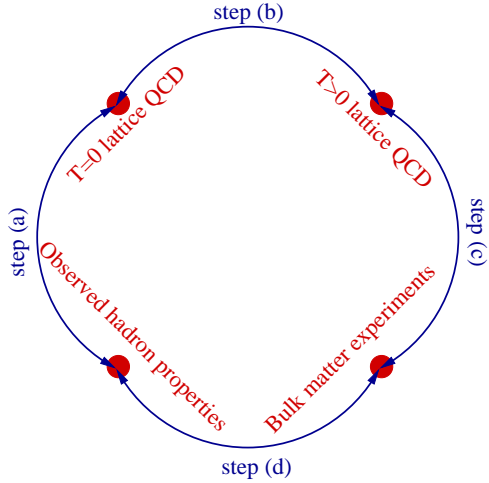}
\end{center}
\caption{Illustration of the chain of reasoning for testing QCD in the non-perturbative domains 
of the strong interactions and obtaining the scale, $\tc$, of the QCD phase diagram.}
\label{fig.cor}\end{figure}

One of the most basic questions to ask about bulk hadronic matter is the value of $\tc$. This 
can be represented as a link 
in a ``circle of reasoning'' that encompasses all the regimes 
of non-perturbative QCD (Fig.1). So far the strategy to find $\tc$ has been 
indirect: first lattice QCD computations are performed at both $T=0$ and $T>0$ in order to determine 
a ratio $\tc/m$, where $m$ is a typical hadronic scale (step (b) of Fig.1).  
Then one replaces the scale $m$, determined on the lattice, by an experimental measurement (step (a)).  
The temperature at each $\snn$ extracted from models of particle yields \cite{peter,cleymans} is 
step (d) of the circle of reasoning. From such models one finds that the fireball of bulk nuclear matter created 
in heavy-ion collisions, which is initially out of equilibrium, evolves to a state of thermal 
equilibrium at chemical freeze-out. The models do no give $\tc$; however, they allow the 
extraction of $T$ and $\mub$ at freezeout. 
In this paper we show that predictions of lattice computations of finite temperature 
QCD \cite{ilgti}, taken in conjunction with determinations of $T_c$ in step (b) \cite{lattice1,lattice2,lattice3,lattice4} agree well with experimental measurements 
on bulk hadronic matter \cite{star}. This allows us to invert the reasoning and extract $\tc$ directly from the experimental
measurements in heavy-ion collisions (step (c)). 
The agreement of the
temperature from steps (c) and (d) along with the agreement of $\tc$ extracted from steps 
(a) and (b) with that from (c) show the complete compatibility of a single theory of hadron 
properties and of bulk QCD matter, \ie, of all non-perturbative regimes of the strong
interactions. This approach may present a new domain of tests of the standard model of particle physics.

\begin{figure}
\begin{center}
\includegraphics[scale=0.55]{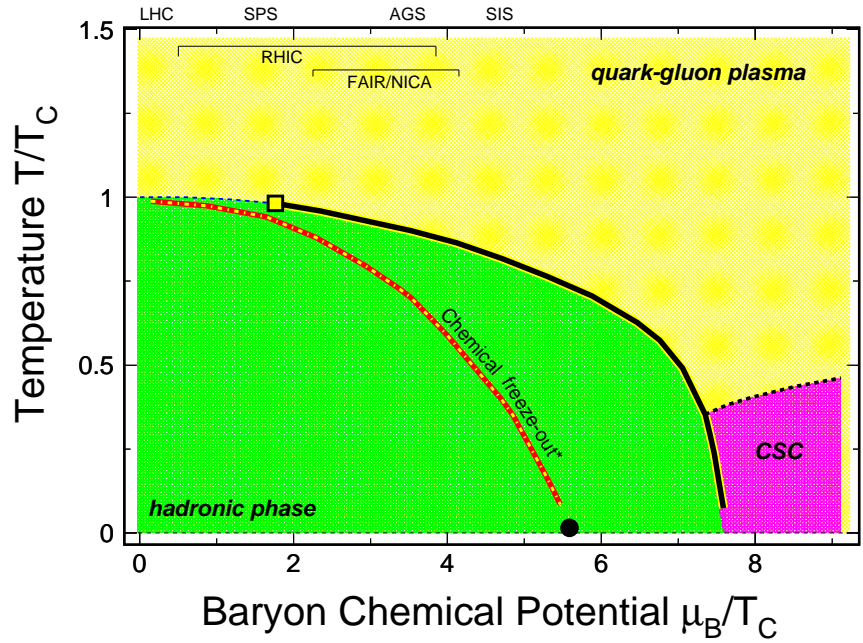}
\end{center}
\caption{Current conjectures for the QCD phase diagram. The phase boundary
(solid line) between the normal low-temperature hadronic phase of bulk
QCD matter and the high-temperature partonic phase is a line
of first order phase transitions which begins at large $\mub$ and small
$T$ and curves towards smaller $\mub$ and larger $T$. This line ends
at the QCD critical point whose probable position, derived from
lattice computations, is marked by a square. At even smaller $\mub$
there are no phase transitions, only a line of cross-overs (shown
by a dashed line).  The red-yellow dotted line corresponds to the chemical
freeze-out line from the evolution of the bulk QCD matter produced in
high energy heavy-ion collisions. The solid point at $T=0$ and $\mub=938$
MeV represents nuclear matter in the ground state. At large $\mub$ and low $T$ 
is the color superconductor phase (CSC)\cite{alford}.}
\label{fig.phased}\end{figure}

\textbf{The conjectured phase diagram of QCD:}
In the current
conjectures for the parts of the phase diagram that is accessible with heavy-ion
collisions \cite{phased} (Fig.2), calculations within simplified models which
mimic QCD show that at large $\mub$ there is a first order hadron-QGP
phase transition. This phase boundary is expected to end in a critical point at finite $\mub$,
as lattice computations \cite{lattice1,lattice2,lattice3,lattice4} agree with general symmetry
arguments \cite{universality}, which indicate that at $\mub=0$ there
is neither a first-order nor a second-order phase transition but only
a cross-over at $\tc$. The determination of $\tc$ sets the scale of
the QCD phase diagram. Current best estimates of the position of the
critical point \cite{latreview} are reflected in the position indicated in Fig.2.  
The experimental focus currently is on an attempt to
locate the critical point and the line of phase coexistence \cite{bm1,bm2}.

By changing $\snn$ one traces out a line of chemical freeze-out in the phase diagram, 
as shown in Fig.2.  This line is parameterized through a hadron resonance gas model \cite{peter,cleymans}.
Because this work focuses on making a connection between QCD thermodynamic
calculations and observables measured in experimental facilities, we
also show in Fig.2 the range of $\mub$ values covered by various
experiments as one traverses the chemical freeze-out line by changing
$\snn$.  The solid point around $\mub$ = 938 MeV is the location of
ordinary nuclear matter \cite{relativistic}, the best characterized
point on the phase diagram.

\textbf{Comparison of experimental measurements with lattice QCD predictions:}
Lattice QCD computations leave open the question of a scale and yield
dimensionless predictions, for example for $P/T^4$ as a function of
$T/T_c$ and $\mub/T$. Here we discuss the non-linear susceptibilities (NLS) of baryons, $\nls{n}$, of order $n$
\cite{nlsdef}. These are the Taylor coefficients in the expansion of $P$
with respect to $\mub$ at fixed $T$ in the usual dimensionless form
\begin{equation}
  T^{n-4} \nls{n}\left(\frac T{T_c},\frac{\mub}T\right) =
        \frac1{T^4}\, \left.\frac{\partial^n}{\partial(\mub/T)^n}\;
              P\left(\frac T{T_c},\frac{\mub}T\right)\right|_{T/T_c}.
\label{nls}\end{equation} 
Lattice measurements of the series expansion of the NLS in powers of $\mub/T$ are resummed using Pade
approximants in order to give predictions for the above quantities \cite{ilgti}. 
They are of interest because they are related to cumulants of the
fluctuations of the baryon number in thermal and chemical equilibrium
in a grand canonical ensemble.

The $n$-th cumulant of such fluctuations, $[B^n]$, is given by
\begin{equation}
   [B^n] = VT^3 T^{n-4} \nls{n}\left(\frac T{T_c},\frac{\mub}T\right),
\label{cumulant}\end{equation}
where $V$ is the volume of the observed part of the fireball.
Because observed hadrons are in thermal and chemical equilibrium at
the freeze-out, this relation should hold for cumulants of the observed
event-by-event distribution of net-baryon number in heavy-ion collisions.
The cumulants are often reported as the variance, $\sigma^2=[B^2]$, the
skewness, $S=[B^3]/[B^2]^{3/2}$ and the Kurtosis, $\kappa=[B^4]/[B^2]^2$.
It is clear from these definitions that the $V$-dependence in Eq. \
\ref{cumulant} gives the correct volume scaling expected from the
central limit theorem. This leads to the classic extraction of the
susceptibility from fluctuations in the grand canonical ensemble \cite{twinprl1,twinprl2}.

There is one remaining subtlety in comparing lattice computations with
experimental data. Most experiments are designed to measure event-by-event
net-protons.  The data discussed in the current work is from the STAR
experiment at RHIC \cite{star}, which identifies protons and anti-protons by measuring
the specific ionization energy loss of these particles in the gas of a
Time Projection Chamber. These measurements miss neutrons,
the other dominant part of the baryon distribution. This may impose 
limitations on our measurement of fluctuations.
However the effect of isospin fluctuations on the shape of the net-baryon distributions is small \cite{hatta}.
Hence we proceed under the
assumption that the shape of the net-proton distributions reflects the
net-baryon distributions up to distortions smaller than the estimated
errors in measurements of the cumulants.

We are unable to exploit Eq. \ \ref{cumulant} directly in
heavy-ion experiments because the volume, $V$, is hard to determine
precisely experimentally.  However, the ratios
\begin{eqnarray}
\nonumber
  (m_1):&&  S\sigma = \frac{[B^3]}{[B^2]} = \frac{T\nls3}{\nls2},\\
\nonumber
  (m_2):&&  \kappa\sigma^2=\frac{[B^4]}{[B^2]} = \frac{T^2\nls4}{\nls2},\\
  (m_3):&&  \frac{\kappa\sigma}S=\frac{[B^4]}{[B^3]} = \frac{T\nls4}{\nls3},
\label{ratios}\end{eqnarray}
do not contain the volume and therefore provide a direct and convenient
comparison of experiment and theory \cite{cpod}. The above equations
are written in a form that emphasizes this connection--- the left hand
side can be measured in an experiment whereas the right hand side can
be predicted by lattice QCD. We use the notation $m_{1,2,3}$ generically
to refer to either side. 
\begin{figure}
\includegraphics[scale=0.85]{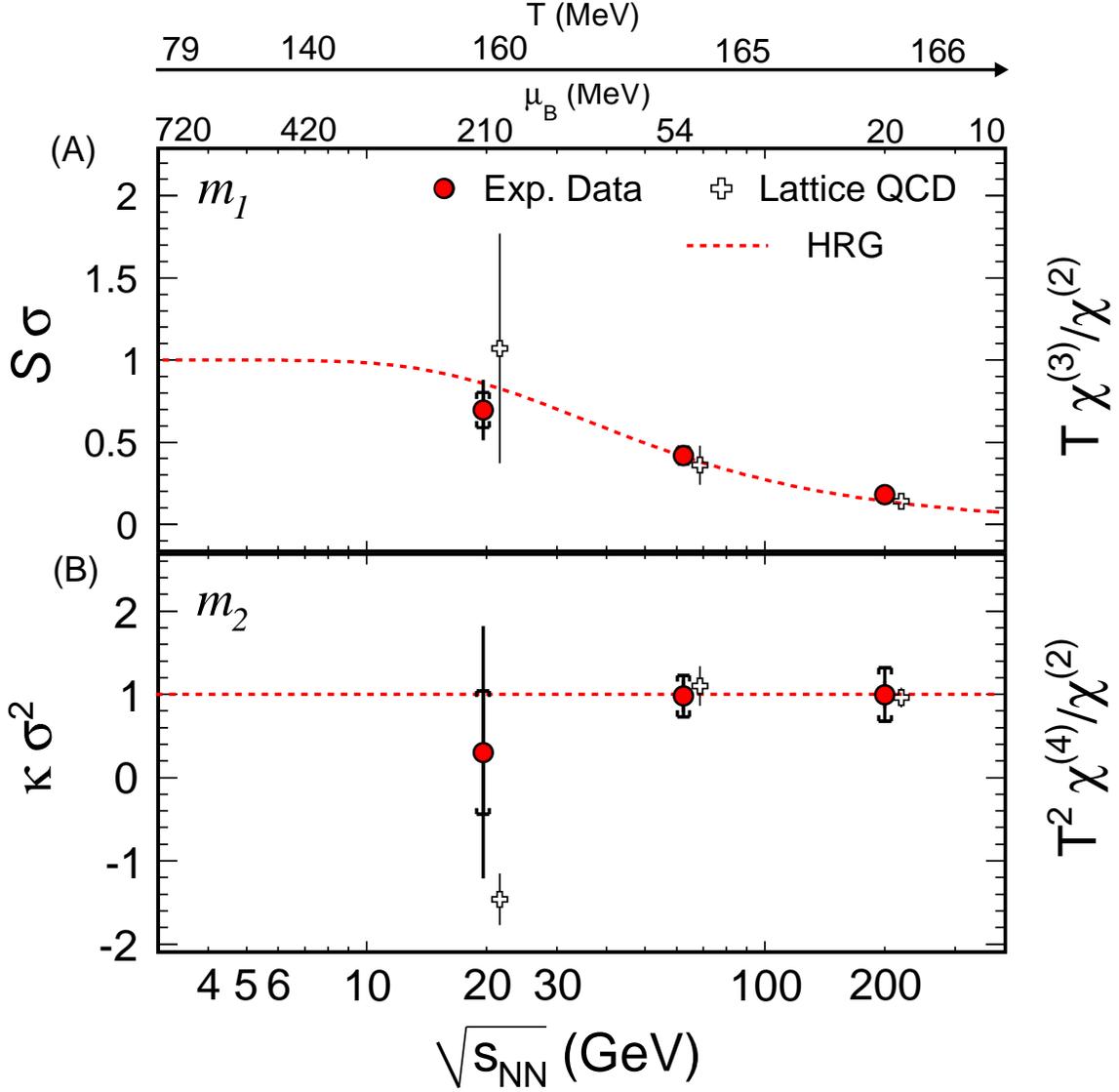}
\caption{Comparison of lattice QCD and experimental data for $m_1$
(A) and $m_2$ (B). Experimentally measured ratios of cumulants of net-proton
distributions, $m_1=S\sigma$ and $m_2=\kappa\sigma^{2}$ are shown as
a function of $\snn$ for impact parameter values of less than 3 fm for
Au+Au collisions at RHIC \cite{star}. Also plotted on the top scale are the chemical freeze-out
values of $\mub$ and $T$ corresponding to $\snn$ as obtained from a
hadron resonance gas model, which assumes the system to be in chemical
and thermal equilibrium at freeze-out \cite{peter,cleymans}. The prediction
of such a model for $m_1$ \cite{karsch} is shown as the dashed red line. The lattice
predictions for these quantities are drawn from a computation with lattice
cutoff of $1/a\simeq960-1000$ MeV and converted to the dimensional scale of $T$
and $\mu$ using $T_c=175$ MeV.}
\label{fig.m1m2}\end{figure}

We now proceed to discuss the comparison of $m_1$ and $m_2$ from
experiment and theory (Fig.3). 
The experimental measurements \cite{star} were made using the
number of protons ($p$) and anti-protons ($\bar{p}$) produced in the collision of Au
ions around $90^{0}$ to the beam axis with the impact parameter of the
collisions being less than 3 fm \cite{glauber}. $p$ and $\bar p$
are in the range of 400 MeV/$c$ to 800 MeV/$c$. This
choice of momentum range is designed to obtain the purest sample of
$p$ and $\bar p$. A large fraction of $p$ and $\bar p$
is contained in this kinematic range. The effect of finite 
reconstruction efficiency of $p$ and $\bar p$  has been shown to be negligible \cite{star}. 
The experimental values of $S\sigma$ and $\kappa\sigma^2$ 
are shown as a function of $\snn$.

The lattice calculations \cite{ilgti} were carried out using
two flavors of staggered quarks in QCD. The lattice cutoff $1/a\simeq960-1000$
MeV and the bare quark mass were tuned to give a pion mass of about 230
MeV \cite{pionmass}. These computations were performed at $\mub=0$ and the Taylor series
coefficients of $P/T^4$ were used to extrapolate $m_1$
and $m_2$ to the freeze-out conditions using appropriate order Pad\'e
approximants to resum the series expansions. Since lattice results are
obtained in terms of $T/T_c$ and $\mub/T$ (see Eq. \ \ref{nls}), their
extrapolation to the freeze-out conditions required the input of $\tc$.
The lattice values were obtained using $\tc=175$ MeV, compatible with
indirect determinations of $T_c$ \cite{lattice1,lattice2,lattice3,lattice4}.

On the upper scales of Fig.3 we also show the $\mub$ and $T$ values 
at chemical freeze-out that correspond to the various $\snn$. For this we used the 
functional relationship between these values from the hadron resonance gas model 
using the yields of hadrons discussed in \cite{peter,cleymans}. The model predictions 
for $m_1$ \cite{karsch} are also shown. The hadron resonance gas model predictions can 
be reproduced if baryon and anti-baryon numbers are independently Poisson distributed. 
Having established a connection between $\snn$ 
and ($T$, $\mub$) we compare the experimental data on fluctuations to those predicted 
from lattice QCD. 
Excellent agreement is
seen between lattice QCD predictions and experimental measurements for
all three beam energies. This marks the first successful direct test
of QCD against experimental data in the non-perturbative context of bulk
hadronic matter.  The agreement with the data is yet another non-trivial
indication that the fireball produced in heavy-ion collisions is in
thermal and chemical equilibrium at chemical freeze-out.

\begin{figure}
\includegraphics[scale=0.85]{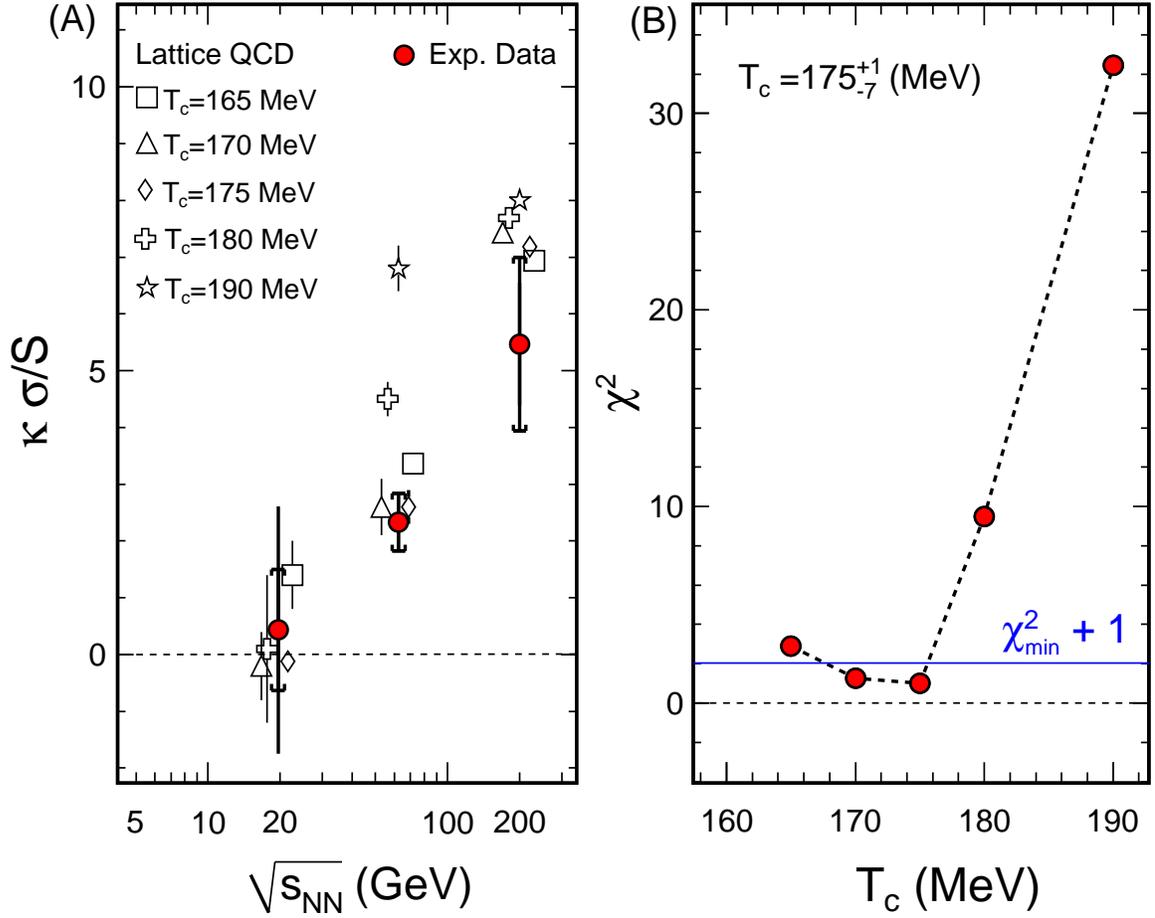}
\caption{Comparison of $m_3$ from experiment and lattice predictions, and
the extraction of $\tc$. (A): $\kappa\sigma/S$ of net-proton
distribution measured in collisions of Au ions at varying
$\snn$ and with an impact parameter of less than 3 fm. This is compared to
lattice QCD predictions with cutoff $1/a\simeq960-1000$ MeV for the corresponding
ratio of susceptibilities extrapolated to the freeze-out conditions using
different values of $\tc$. The lattice results at each $\snn$ are slightly
shifted for clarity in presentation. (B): The comparison of experimental data
and lattice QCD predictions, shown through $\chi^2$ as a function of
$\tc$ using the definition given in Eq. \ref{fit}. This yields the estimate
of $\tc$ and its errors as discussed in the text.}
\label{fig.tc}\end{figure}

\textbf{Setting the scale of bulk QCD:} 
Lattice QCD results for $m_{1,2,3}$ are obtained for dimensionless
arguments $T/\tc$ and $\mub/T$ as shown in Eq.\ 2. For a given
value of $\snn$, the experimental observations are realized at the
corresponding chemical freeze-out, characterized by $T$ and $\mub$.
Thus a comparison of the two requires a choice of the scale, $\tc$. 
By varying this scale to obtain the best fit between the QCD
predictions and experimental measurements, we determine $\tc$ 
for the first time through an observable connected to strongly 
interacting bulk matter. The results are, of course, subject to 
all the caveats expressed in the previous section. The observable 
that we choose for comparison is $m_3$. The lattice computation of this 
quantity has the smallest systematic uncertainties among the three 
explored here, and thus is the best quantity to use to constrain $\tc$.

Figure 4A shows the comparison of $m_3$
between experimental results from Au ion collisions and lattice QCD
predictions. This is an extension of Fig.\ 3 which shows
a comparison with $m_1$ and $m_2$. In this analysis, the results of  
$m_1$, $m_2$ and $m_3$ are consistent as required in Eq.\ \ref{ratios}.
The new information here is that we
show lattice predictions obtained with different values of $\tc$.
The errors on the experimental data points are statistical (lines) and
systematic (brackets) errors \cite{star}.  
The errors bars on the lattice
predictions are statistical errors with a
cutoff of $1/a\simeq960-1000$ MeV. The lattice spacing effects and the
effect of tuning the bare quark mass are the main sources of remaining
uncertainties in the predictions. These are not parameterized as systematic
uncertainties. However, it is known that their effect is small at the
two highest values of $\snn$ \cite{ilgti}. 

In order to arrive at a quantitative estimate of the scale parameter
$\tc$ we perform a standard statistical analysis. For each value of $\tc$
we compute,
\begin{equation}
  \chi^2(\tc) = \sum_{\snn}
    \frac{[m_3^{\mathrm{expt}}(\snn) - m_3^{\mathrm{QCD}}(\snn,\tc)]^2}
       {\mathrm{Error}^2_{\mathrm{expt}} + \mathrm{Error}^2_{\mathrm{QCD}}}
\label{fit}\end{equation}
where the errors in the experimental and lattice QCD quantities are
obtained as explained above. 
The lattice predictions are calculated for the grid of $\tc$ values
(Fig.\ 4). The minimum of
$\chi^2$, corresponding to the most probable value of the parameter
being estimated, occurs at $\tc=175$ MeV.  The standard errors on the
parameter are the values of $\tc$ for which $\chi^2$ exceeds the minimum
value by unity. It is clear from Fig.4(B) that
this is bounded by $+5$ and $-10$ MeV.  A piece-wise linear interpolation
between the grid points yields the more reliable error estimate,
$+1$ and $-7$ MeV. By comparing different interpolation schemes we
find that the error estimate is stable. As a result we conclude that
\begin{equation}
   \tc = 175^{+1}_{-7} \;\;{\mathrm MeV}.
\label{tcval}\end{equation}
The error estimates include systematic and statistical errors on experimental
data but only statistical errors on the lattice QCD computations.

The result in Eq. \ \ref{tcval} is compatible with current indirect
estimates of $\tc$ which come from setting the scale of thermal lattice
QCD computations via hadronic observables.  Furthermore, this gives a
scale for temperatures which is compatible with the resonance gas model,
as shown in Fig.3. As we discussed in the introduction,
this closes a circle of inferences which shows that phenomena obtained in
heavy-ion collisions are fully compatible with hadron phenomenology, and
provides a first check in bulk hot and dense matter for the standard model
of particle physics.

\textbf{Conclusions and Outlook:} 
We have performed a direct comparison between experimental data from high
energy heavy-ion collisions on net-proton number distributions and lattice
QCD calculations of net-baryon number susceptibilities.  The agreement
between experimental data, lattice calculations and
a hadron resonance gas model indicates that the system produced in
heavy-ion collisions attained thermalization during its evolution.
The comparison further enables us to set the scale for non-perturbative
high temperature lattice QCD by determining the critical temperature
for the QCD phase transition to be $175^{+1}_{-7}$ MeV.

This work reveals the rich possibilities that exist for a
comparative study between theory and experiment of QCD thermodynamics
and phase structure.  In particular, the current work can be extended
to the search for a critical point.  In a thermal system, the correlation length
($\xi$) diverges at the critical point. $\xi$ is related to various moments of the
distributions of conserved quantities such as net-baryons, net-charge, and
net-strangeness. Finite size and dynamical effects in heavy-ion collisions
put constraints on the values of $\xi$ \cite{krishnaxi}.  The lattice calculations
discussed here and several QCD-based models have shown that moments of
net-baryon distributions are related to  baryon number susceptibilities
and that the ratio of cumulants $m_2=\kappa\sigma^2$, which is related to the
ratio of fourth order to second order susceptibilities, shows a  large
deviation from unity near the critical point.  Experimentally,  $\kappa\sigma^2$
can be measured as a function of $\snn$ (or $T$ and $\mub$) in heavy-ion
collisions. A non-monotonic variation of  $\kappa\sigma^2 $ as a function
of $\snn$ would indicate that the system has evolved in the vicinity of
the critical point and thus could be taken as evidence for the existence of a critical
point in the QCD phase diagram.

\bibliography{scibib}

\begin{thebibliography}{99}
\bibitem{relativistic}
   Note that $\mub$ of bulk nuclear matter is quoted in the usual convention
   adopted in relativistic treatments, since anti-baryon production also needs
   to be accounted for. An alternative (non-relativistic) definitions which
   takes nucleon number to be fixed would give much smaller chemical potential.
   However, this does not correspond to the physics of baryon number
   fluctuations which we examine.
\bibitem{glauber}
   The impact parameter is determined through a Glauber Monte Carlo
   procedure. The selected events correspond to the most central
   0-5\% and 0-10\% events of the total hadronic cross section in Au+Au collisions for
   $\snn$ = 200, 62.4 and 19.6 GeV respectively.
\bibitem{pionmass}
The change in the radius of convergence in going from
$m_\pi/m_\rho=0.33$ to $0.2$ is likely to be between 10--15\%
(22). The corresponding effect on $m_3$ is about 2\% or less at the two
highest energies, and less than 20\% at energy of 19.6 GeV.
\end{thebibliography}
\bibliographystyle{Science}

\begin{scilastnote}
\item 
We thank R. V. Gavai, F. Karsch, D. Keane, V. Koch, B. Mueller, K. Rajagopal, K. Redlich, H. Satz, 
M. Stephanov for enlightening discussions. We acknowledge the Indian 
Lattice Gauge Theory Initiative for computational support, 
DAE-BRNS through the project sanction No. 2010/21/15-BRNS/2026, 
US Department of Energy under Contract No. DE-AC03-76SF00098 and
the Chinese Ministry of Education.
\end{scilastnote}

\author{
}

This manuscript has been accepted for publication in Science. 
This version has not undergone final editing. Please refer to the complete version of 
record at http://www.sciencemag.org/. The manuscript may not be reproduced or used in any 
manner that does not fall within the fair use provisions of the Copyright Act without the prior, 
written permission of AAAS.

\end{document}